\documentstyle[aps,preprint,epsf]{revtex}
\begin{document}
\draft
\title{The Effect of Anisotropy on Vortex Lattice Structure \\
and Flux Flow in d-Wave Superconductors}
\date{\today}
\author{
D. Chang\footnote{Department of Physics,
National Tsing Hua University,
Hsinchu, Taiwan, 30043, R.O.C.},
C.-Y. Mou$^*$,
B. Rosenstein\footnote{Electrophysics Department,
National Chiao Tung University,
Hsinchu, Taiwan, 30043, R.O.C.},
and C. L. Wu$^*$
}
\maketitle

\begin{abstract}
We describe effects of anisotropy caused by the crystal lattice in d-wave
superconductors using effective free energy approach in which only one order
parameter, the d-wave order parameter field, is used. All the effects of
rotational symmetry breaking, including that of the s-wave mixing, can be
parametrized by a single four derivative term. We find solutions for single
vortex and the vortex lattice. Extending the formalism to include the time
dependence, effects of anisotropy on moving vortex structure are calculated.
Both direct and Hall I-V curves as functions of the angle between the
current and the crystal lattice orientation are obtained.
\end{abstract}


%\author{
%D. Chang,
%\footnote{Department of Physics,
%National Tsing--Hua University,
%Hsinchu, Taiwan, 30043, R.O.C.}
%C.-Y. Mou$^*$,
%B. Rosenstein
%\footnote{
%Electrophysics Department,
%National Chiao Tung University,
%Hsinchu, Taiwan, 30043, R.O.C.}
%and C. L. Wu$^*$
%}

It is widely believed that the superconductivity in layered high $T_{c}$
cuprates is largely due to the $d_{(x^{2}-y^{2})}$ pairing \cite{Annett}. It
has been observed that this influences the vortex lattice structure \cite
{Keimer,Maggio}. There are also indications that even though the major bulk
pairing mechanism is of $d$ wave nature, there is a small admixture of the
s-wave pairs in the condensate. Ren et al \cite{Ting1}, using a
phenomenological microscopic model (in the weak-coupling limit), and
Soininen et al \cite{Berlinsky1}, considering attractive nearest neighbors
interaction, proposed an effective Ginzburg-Landau (GL) type theory using
two fields. Two order parameters, $s$ and $d$, describe the gap functions in
corresponding channels. The free energy was constructed to include the
fourfold symmetry $D_{4h}$\cite{Berlinsky2}:

\begin{eqnarray}
f &=&\alpha _{s}|s|^{2}-\alpha _{d}|d|^{2}+\beta _{1}|s|^{4}+\beta
_{2}|d|^{4}+\beta _{3}|s|^{2}|d|^{2}+\beta _{4}(s^{*2}d^{2}+d^{*2}s^{2}) 
\nonumber \\
&&+\gamma _{s}|{\bf \Pi }s|^{2}+\gamma _{d}|{\bf \Pi }d|^{2}+\gamma
_{v}[s^{*}(\Pi _{y}^{2}-\Pi _{x}^{2})d+c.c.]  \label{twofield}
\end{eqnarray}
where ${\bf \Pi }\equiv -i{\bf \nabla }-e^{*}{\bf A}$ (with $c=\hbar =1$).
Using equations following from this free energy or using more fundamental
equations \cite{Ichioka}, one obtains a characteristic four-lobe structure
for the s-wave inside a single vortex \cite{Berlinsky2,Blatter,Ting3}. The
distribution of the magnetic field was also obtained recently \cite
{Blatter,Berlinsky2}. The s-wave vanishes outside the core, while the d-wave%
{\it \ }distribution becomes rotationally invariant, indistinguishable from
the usual Abrikosov solution. Since the fourfold vortex core structure is
seemingly in conflict with the symmetry of the triangular lattice, the
asymmetry of vortices therefore can, in principle, distort the usual
triangular vortex lattice. If one were to look for differences in the
behavior of vortices between this case and that of the conventional s-wave
superconductors, one would like to be closer to $H_{c2},$ so that the
influence of the core will be more significant. Another phenomenon in which
the core plays a major role is the dissipation in the course of the flux
flow.

In this paper we study the above two phenomena using time independent and
time dependent effective GL equations within a simplified {\it one component
framework}. The simplicity of the formulation allows us to clarify,
essentially without loss of generality and the use of numerical methods,
various delicate questions about single vortex and the vortex lattice. The
degrees of freedom we include in the analysis contain: (1) an arbitrary
rotation angle $\varphi $ between the crystal lattice and the vortex
lattice\ and (2) all the possible lattices, not only the rectangular ones
considered before \cite{Berlinsky1,Berlinsky2,Ting3}). This is the first
time that the lattice is\ demonstrated to be centered rectangular ({\it CR})
with the most general lattice included in the analysis. Moreover the
treatment can be extended to moving flux lattices, which, as is well known 
\cite{Dorsey,Hu} are more demanding, as far as calculational complexity is
concerned. \ 

Within the two-field formulation, Soininen at al \cite{Berlinsky2} observed
that, in a predominantly d-wave superconductor, the s-wave component is
generally very small: it is ''induced'' by the variations of the larger $d$
component. In the bulk, only the $d$ field acquires a nonzero value, while,
near the core, the rotationally noninvariant gradient term $s^{*}(\Pi
_{y}^{2}-\Pi _{x}^{2})d$ ''communicates'' the deviations from the condensate
value of $d$ to that of $s$. Even near the core, where $d$ is small, it is
still larger then $s$ by a factor of 20. Since the $s$ field is ''induced''
by $d$, it varies on the scale of the d-wave coherence length $\xi _{d}=%
\sqrt{\gamma _{d}/\alpha _{d}}$. Then $\gamma _{s}\Pi ^{2}s\sim \left(
\gamma _{s}/\xi _{d}^{2}\right) s$ is small compared to $\alpha _{s}s$ if $%
(\gamma _{s}/\gamma _{d})(\alpha _{d}/\alpha _{s})\ll 1$ (typically $\gamma
_{s}/\gamma _{d}\sim 1$). This is an excellent approximation in both near
and far regions from the core \cite{Berlinsky2,Ting2,Blatter}. Therefore,
the field $s$ is, to the first order in $1/\alpha _{s}$, $s=-(\gamma
_{v}/\alpha _{s})(\Pi _{y}^{2}-\Pi _{x}^{2})d.$ Substituting this back to GL
equations, we obtain, to first order in $1/\alpha _{s}$, the effective free
energy 
\begin{equation}
f_{eff}[d]=\frac{1}{2m_{d}}|\Pi d|^{2}-\alpha _{d}|d|^{2}+\beta |d|^{4}-\eta
d^{*}\left( \Pi _{y}^{2}-\Pi _{x}^{2}\right) ^{2}d.  \label{fe}
\end{equation}
Here we have replaced $\gamma _{d}$ by a more conventional notation $%
1/2m_{d} $ and defined $\eta \equiv $ $\gamma _{v}^{2}/\alpha _{s}$. The
last term should be treated as a perturbation.

Note that the two-field equations are highly nontrivial even at the
linearized level, so that authors of Ref. \cite{Berlinsky2} resort to the
variational estimate in order to find a solution. The linearized one-field
equation can be solved perturbatively in $\eta $. The advantage of this
equation, especially as far as relation to experiments is concerned, is that
the number of coefficients is much smaller: instead of 10 parameters in the
two-field free energy, there is just 3. One can further motivate the use of
the effective free energy Eq.(\ref{fe}) even with no connection to the
two-field formalism. Generally superconductivity is a phenomenon of
''spontaneous gauge symmetry breaking '' phenomenon irrespective of the
mechanism of pairing or channels in which it occurs. The symmetry should be
represented by a single order parameter: the superconducting phase. While in
pure s-wave or d-wave superconductors the phase is simply identified with
the phase of the gap function, in more complicated microscopic theories with
few channels opened, the superconducting phase is just the common phase of
various gap functions. Therefore, quantities other than the phase which
enter various phenomenological GL type equations, although useful, are not
directly related to the spontaneous gauge symmetry breaking.

The general form of the effective free energy can be obtained just by the
dimensional analysis and symmetry. It is well known that in a $D_{4h}$
symmetric field theory, at the level of the dimension three (relevant)
terms, full rotational symmetry is restored. Therefore, one has to consider
''irrelevant'' dimension five terms in order to break the rotational
symmetry down to $D_{4h}$. $d^{*}(\Pi _{y}^{2}-\Pi _{x}^{2})^{2}d$ is the
only such term which does not respect rotational symmetry and therefore is
important for studying anisotropy effects. Consequently only one coefficient 
$\eta $ is needed to describe the anisotropy effects. The origin of $\eta $
might, in principle, come from sources other than the d-s mixing. The one
field formulation also avoids the problem of the artificial second phase
transition at $T_{s}$ assuming $\alpha _{s}=\alpha ^{\prime }(T_{s}-T)$ in
the two-field formulation.

{\it Single vortex solution near }$H_{c1}${\it .} In strongly type II
materials, we can safely ignore the magnetic field. The GL equation can be
solved perturbatively as follows: Write $d=d_{0}+\lambda d_{1}$ , where $%
d_{0}=f_{0}(r)e^{i\phi }$ is the solution of the standard unperturbed GL
equation and $\lambda \equiv 4\eta \,m_{d}^{2}\alpha _{d}$ is a
dimensionless small parameter. The angular dependence of $d_{1}$ is easily
found to contain only three harmonics $e^{-3i\phi }$, $e^{+i\phi }$ and $%
e^{5i\phi }$ 
\begin{equation}
d_{1}(r,\phi )=f_{-3}(r)e^{-3i\phi }+f_{1}(r)e^{i\phi }+f_{5}(r)e^{5i\phi }.
\end{equation}
This is consistent with the fourfold symmetry which is built into the
theory. The analytic expression for $f_{0}$ does not exist, \ but there are
a number of known approximations. Using one of them, $f_{0}(r)=r/\sqrt{%
r^{2}+\xi _{v}^{2}}$ , the set of linear equations for $f_{1},f_{2},f_{3}$
are then solved numerically (the $f_{3}$ equation decouples from the other
two) \cite{paper}. The d-wave configuration is nearly indistinguishable from
that of the two-field formalism as expected, while the s-wave configuration
calculated with $s=-(\gamma _{v}/\alpha _{s})(\Pi _{y}^{2}-\Pi _{x}^{2})d$
has the four-lobe structure with four zeroes on the axes (same as in \cite
{Berlinsky2} but different from \cite{Ting2}).

{\it Vortex lattice. }In the opposite limit near $H_{c2}$ one first neglects
the nonlinear terms in the GL equation and finds the lowest energy solutions 
$\Psi _{k_{n}}(x,y)$ of the linearized equation. The lattice solution is
constructed from the linear superposition of $\Psi _{k_{n}}(x,y)$ in such a
way that it is invariant under the corresponding symmetry group of the
lattice: $d(x,y)=\sum_{n}C_{n}\Psi _{k_{n}}(x,y)$. A general lattice in 2D
can be specified by three parameters $a,b$ and $\alpha $, where $a$ and $b$
are the two lattice constants, while $\alpha $ is the angle between the two
primitive lattice vectors as defined in Ref. \cite{SaintJames}. Flux
quantization provides a constraint: $Hab\sin \alpha =\Phi _{0}$. In the
d-wave superconductors the rotational symmetry is broken, therefore the
relative orientation of the vortex lattice to the underlying lattice becomes
important. We denote $\varphi $ to be the angle between $\stackrel{%
\rightarrow }{a}$ and one of the axes of the underlying lattice. The most
general lattice is built into the solution via appropriate relations among $%
C_{n}$\cite{SaintJames}. The lattice structure is then found by minimizing
the free energy or equivalently, the Abrikosov's parameter $\beta _{A}\equiv
\langle |d|^{4}\rangle /\langle |d|^{2}\rangle ^{2}$. Here $\langle
...\rangle $ is the average over space.

We solve the linearized GL equation perturbatively in the anisotropy
parameter $\eta $ in the Landau gauge, ${\bf A}=(0,Hx)$(see details in \cite
{paper}): 
\begin{eqnarray}
\psi (x,y) &=&\left( \frac{1}{\pi l_{H}^{2}}\right) ^{1/4}\exp \left(
iyk_{n}\right) \left[ 1+\eta m_{d}e^{*}H\frac{e^{4i\varphi }}{16}H_{4}\left( 
\frac{x}{l_{H}}-k_{n}k_{H}^{2}\right) \right]  \label{perturbedd} \\
&&\times \exp \left[ -\frac{1}{2}\left( \frac{x}{l_{H}}-k_{n}l_{H}^{2}%
\right) ^{2}\right] ,
\end{eqnarray}
where $H_{4}$ is the fourth Hermit polynomial and $l_{H}\equiv \sqrt{1/e^{*}H%
}$. The upper critical field is:

\begin{equation}
H_{c2}(T)=\frac{2m_{d}\alpha ^{\prime }}{e^{*}}\left[ \left( T_{c}-T\right)
+8\eta m_{d}^{2}\alpha ^{\prime }\left( T_{c}-T\right) ^{2}\right] .
\label{hc2t}
\end{equation}
Note that $H_{c2}(T)$ bends upwards around $T_{c}$, in agreement with the
two-field results \cite{Berlinsky2}. This effect has been reported in some
experiments. However one should be cautioned against taking this too
seriously. As we have discussed, there is another rotationally invariant
term of the same dimensionality $\tau $$d^{*}(\Pi ^{2})^{2}d$ which gives
contribution similar to Eq.(\ref{hc2t}): $\Delta H_{c2}(T)=-\tau \frac{%
2m_{d}\alpha ^{\prime }}{e^{*}}\left( T_{c}-T\right) ^{2}.$ This means that
for large enough positive $\tau $ the sign of the $H_{c2}(T)$curvature
changes.

To present $\beta _{A}$ it is convenient to use variables: $\rho +i\sigma $ $%
\equiv \zeta \equiv \frac{b}{a}\exp (i\alpha )$ \cite{SaintJames}. After
some rather nontrivial calculations we obtain, $\beta _{A}=\beta
_{A}^{0}+\eta ^{\prime }\beta _{A}^{1}$ with 
\begin{equation}
\beta _{A}^{0}=\sqrt{\sigma }\left\{ \left| \sum_{n=-\infty }^{\infty
}w_{n}(\zeta )\right| ^{2}+\left| \sum_{n=-\infty }^{\infty }w_{n+1/2}(\zeta
)\right| ^{2}\right\} .  \label{beta0}
\end{equation}
\[
\beta _{A}^{1}=\frac{1}{4}\sqrt{\sigma }%
%TCIMACRO{\func{Re}}
%BeginExpansion
\mathop{\rm Re}%
%EndExpansion
\left\{ e^{4i\varphi }\left[ \sum_{n^{\prime }}w_{n^{\prime }}^{*}(\zeta
)\right] \left[ \sum_{n}w_{n}(\zeta )G_{n}(\sigma )\right] +\left(
n\rightarrow n+\frac{1}{2},n^{\prime }\rightarrow n^{\prime }+\frac{1}{2}%
\right) \right\} 
\]
where $\eta ^{\prime }\equiv \eta m_{d}e^{*}H$ is the dimensionless
parameter appropriate for this situation, $w_{n}(\zeta )=\exp (2\pi i\zeta
n^{2})$ and $G_{n}(\sigma )\equiv (64\pi ^{2}\sigma ^{2}n^{4}-48\pi \sigma
n^{2}+3).$ Having calculated the Abrikosov parameter $\beta _{A}$, one finds
the vortex structure by minimizing it with respect to $\varphi $, $\rho $,
and $\sigma $. The minimization with respect to the angle $\varphi $ is
easily done analytically, while further minimization of $\beta _{A}^{\min
}(\rho ,\sigma )$ is done numerically. The angle $\alpha $ as function of $%
\eta ^{\prime }$ is given on Fig.1. The corresponding $\varphi $ is zero.
For $\alpha =$ $74^{o}$ one gets $\eta ^{\prime }\simeq 0.02.$ There is a
phase transition from {\it CR} to square lattice at $\eta _{c}^{\prime
}=.0235$. Note that this calculation, unlike that for the single vortex, is
valid for arbitrary $\kappa .$ Using standard methods, one can take into
account variations of the magnetic field and calculate corrections to the
magnetization curve using the $\beta _{A}$ calculated here \cite{SaintJames}.

{\it Moving lattice. }The time dependent GL equation describing the time
evolution of the order parameter the one-field formulation is 
\begin{equation}
\gamma \left( \frac{\partial }{\partial t}+ie^{*}\Phi \right) d=-\left( 
\frac{1}{2m_{d}}\Pi ^{2}-\alpha _{d}\right) d+\eta (\Pi _{y}^{2}-\Pi
_{x}^{2})^{2}d-2\beta |d|^{2}d,  \label{TDGL}
\end{equation}
where $\Phi $ is the electric potential. It involves just one additional
parameter $\gamma $ compared to the 2$\times $2 matrix for the two-field
formalism. Although, in principle, this parameter, describing various
dissipation effects, can have a complex part \cite{Dorsey}, we will consider
only real values. To generalize the above procedure for finding the
structure for a moving vortex lattice near $H_{c2}$, one considers the
motion caused by an electric field ${\bf E}$ making an angle $\Theta $ with
the crystal [1,0,0] axis. The vortex lattice velocity is: ${\bf E}=-{\bf v}%
\times {\bf B}.$ For a general direction of the electric field the fourfold
symmetry of the system is completely (explicitly) broken.

Even for the simple s-wave the problem of finding the moving lattice
solution is nontrivial. However in that case there exists the ``Galilean
boost'' trick \cite{Hu} to solve the linearized problem, which is not
applicable to the d-wave. Our approach is to use perturbation theory in $%
\eta ^{\prime }$ to find a complete set of solutions of the linearized
equations and then impose the periodicity conditions to construct the vortex
lattice. It is more convenient to perform the first step in the gauge
aligned in the direction of the electric field, in order to make both the
scalar and vector potentials independent of $y$ and $t$. For the second step
however, it is preferable to use a gauge aligned in the direction of the
vortex lattice. However in general the vortex lattice will not be periodic
along this special direction. To construct this general periodic solution,
one has to solve a periodicity constraint equation for the coefficients $%
C_{k},$ where $k$ is now a continuous index. We then combined the two steps
using gauge transformation \cite{paper}. The results are as follows. The
upper critical field of course depends on velocity

\begin{equation}
H_{c2}(T)=h_{0}+h_{1}(T_{c}-T)+h_{2}(T_{c}-T)^{2}  \label{line}
\end{equation}
with $h_{0}=(m_{d}^{2}/e^{*})\left[ -1+\left( 9+\cos 4\Theta \right) \eta
m_{d}^{3}\gamma ^{2}v^{2}\right] \gamma ^{2}v^{2}$, $h_{1}=2\alpha ^{\prime
}(m_{d}/e^{*})\left( 1-12\eta m_{d}^{3}\gamma ^{2}v^{2}\right) $, $%
h_{2}=16\alpha ^{\prime 2}\eta m_{d}^{3}/e^{*}$. Note that the curvature $%
h_{2\text{ }}$hasn't changed compared to the static case, but we have two
new effects. First of all, the electric field (or, equivalently, electric
current) shifts $H_{c2}$ by a negative constant (proportional to $E^{2\text{ 
}}$) to a lower value. It simply depends on the angle . This is expected.
Secondly, although the curvature $h_{2}$ doesn't change compared with the
static case, the slope $h_{1}$ acquires a negative contribution proportional
to $E^{2}$. In the s - wave case the phase boundary was first found and
discussed in Ref.\cite{Hu}. There are a couple of peculiarities associated
with it like the existence of a metastable normal state and the unstable
superconductive state. The same applies to the present case. As far as we
know, these peculiarities haven't been directly observed in low $T_{c}$
materials. It would be interesting to reconsider this question for the high $%
T_{c}$ materials.

After a lengthy and nontrivial calculation \cite{paper}, the vortex lattice
solution is amazingly simple:

\begin{eqnarray}
\psi (x,y) &=&\sum_{n=-\infty }^{\infty }C_{n}\frac{1}{\sqrt{L}}\left( \frac{%
H}{\pi }\right) ^{1/4}\exp \left( \,ik_{n}y\right) \exp \left[ -\frac{1}{%
2l_{H}^{2}}\left( x-k_{n}l_{H}^{2}\right) ^{2}\right] \times   \nonumber \\
&&\left[ 1+\eta ^{\prime }\sum_{m=1}^{4}c_{m}\frac{e^{im\left( \Theta
+\varphi \right) }}{\sqrt{2^{m}m!}}H_{m}\left( \frac{x}{l_{H}}%
-k_{n}l_{H}-ige^{-i\left( \Theta +\varphi \right) }\right) \right] 
\end{eqnarray}
with $c_{1}=-\sqrt{2}ig\left[ \left( 1+e^{-4i\Theta }\right) g^{2}-2\right]
,c_{2}=-\frac{\sqrt{2}}{2}\left( 1+3e^{-4i\Theta }\right) g,c_{3}=\frac{4%
\sqrt{3}}{3}ige^{-4i\Theta },$ $c_{4}=\frac{\sqrt{6}}{2}e^{-4i\Theta }$ and $%
g\equiv \gamma v.$ The standard Abrikosov's procedure to develop an
approximation for small order parameter around $H_{c2}$ can be applied also
in the flux flow case (see \cite{Hu}). Using this expression the correction
term in the expansion of the Abrikosov parameter $\beta _{A}$ changes and
now the function $G_{n}(\sigma )$in Eq.(\ref{beta0}) depends in a very
simple way on electric field: $G_{n}(\sigma )=e^{4i\varphi }(64\pi
^{2}\sigma ^{2}n^{4}-48\pi \sigma n^{2}+3)-8e^{2i\varphi }g^{2}\cos 2\Theta
(8\pi \sigma n^{2}-1).$ One immediately observes a surprising fact - the
dependence on the angle $\Theta $ and velocity $g$ is only via the
combination $g^{2}\cos 2\Theta $. For example, the resulting lattice for $%
\Theta =\pi /4$ and arbitrary $g$ will be the same as without electric field
at all! Also apparent complete breaking of the rotational symmetry by
general direction of the electric field is not felt by $\beta _{A}.$ Indeed,
the lattice for some arbitrary $\Theta $ and $g^{2}$ is the same as for $%
\Theta =0$ and $g^{2\prime }=g^{2}\cos 2\Theta $. However, the fourfold
symmetry has been reduced. For fixed $\eta ^{\prime }$ and $g^{2}\cos
2\Theta ,$ the minimization was performed numerically and we again obtain
only {\it CR} lattices. The angle $\alpha $ turns out only weakly depend on
the combination $g^{2}\cos 2\Theta $ .

\ This is the first time that the lattice is\ demonstrated to be {\it CR }
type using the most general lattice in the analysis. It turns out that
general oblique lattices have higher energy than the {\it CR }ones despite
the fact that rotational symmetry is completely broken by both the electric
field and by the underlying atomic lattice. Intuitively in the symmetric
case this fact is understandable: {\it CR} lattices are more symmetric,
however for rotationally nonsymmetric or moving lattices this is no longer
so.

{\it I - V curves for the flux flow. }Now we consider the dissipation in
vortex cores due to flux flow. As it is well known, the fourfold symmetry
forces the conductivity tensor $\sigma _{ij},$defined by $J_{i}=\sigma
_{ij}E_{j}$, to be rotationally symmetric. Namely, $\sigma _{ij}=\sigma
\delta _{ij}+\sigma ^{H}\varepsilon _{ij}.$ Here $\sigma $ is the usual
(Ohmic) conductivity, $\sigma ^{H}$ is the Hall conductivity and $%
\varepsilon _{ij}$ is the antisymmetric tensor. The additional term in the
free energy corrects the values of $\sigma $ and $\sigma ^{H},$ but the
correction is of the order $\eta $ and therefore small. So, to see
anisotropy, we definitely would like to go beyond linear response. This has
been done for simple s - wave TDGL \cite{Hu} near $H_{c2\text{ }}$. We will
neglect pinning and consider motion of a very large bundle. While there is a
normal component of the conductivity, here we will concentrate on the
contribution of the supercurrent only. For the discussion of the relative
contribution of the two see \cite{Hu}.

We therefore calculate the current as a function of ${\bf E}$ beyond linear
response. Note that in this case the condensate has to be properly
normalized: $<d^{*}d>=\alpha _{d}/(2\beta \beta _{A})$.  Using the
normalized $d$, we found the anisotropic direct and  Hall currents

\begin{eqnarray}
\Delta {\bf J}_{dir} &=&\eta \frac{2m_{d}\gamma ^{3}E^{3}}{\beta
_{A}^{0}e^{*}H^{4}}\left( 1+\cos 4\Theta \right) ,  \label{direct} \\
\Delta {\bf J}_{Hall} &=&-\eta \frac{2m_{d}\gamma ^{3}E^{3}}{\beta
_{A}^{0}e^{*}H^{4}}\sin 4\Theta .  \label{Hall}
\end{eqnarray}
Note that both direct and Hall currents depend only on the fourth harmonics
of the angle between ${\bf E}$ and the crystal lattice orientation. Only the
cubic power of ${\bf E}$ contributes, all the higher orders terms are
cancelled.

To summarize, we studied the static and dynamical anisotropy effects in
d-wave superconductors using the effective one component model in which
these effects are parametrized by a single parameter $\eta $. For static
lattice we clarified several issues and found the critical value of $\eta
_{c}^{\prime }$ $=.0235$ at which thermodynamic transition to the square
lattice occurs. (The sample of \cite{Keimer} is found to be in CR phase
while that of \cite{Yethiraj} is in square phase). The results of our study
of the moving lattices for the direct current, Hall current magnetization
and phase boundary are very simple and are given in Eqs.\ref{direct},\ref
{Hall},\ref{beta0} and \ref{line} respectively. It would be interesting to
determine $\eta ^{\prime }$ of the same sample from a few of the above
effects.

Authors are very grateful to C.C. Chi, Y.S. Guo, M.K. Wu, V. Yang, S.Y.
Hsu, C.R. Hu, C.S. Ting and F.C. Zhang for discussions and correspondence.
The work is supported by NSC of ROC.

\clearpage

\noindent {\bf Figure Caption}: \newline
Fig. 1 The angle $\alpha $ of the centered rectangular lattice as a function
of the parameter $\eta ^{\prime }$ describing the asymmetry. There is a
phase transition from centered rectangular to square lattice at $\eta _{c}^{^{\prime
}}=0.0235$.

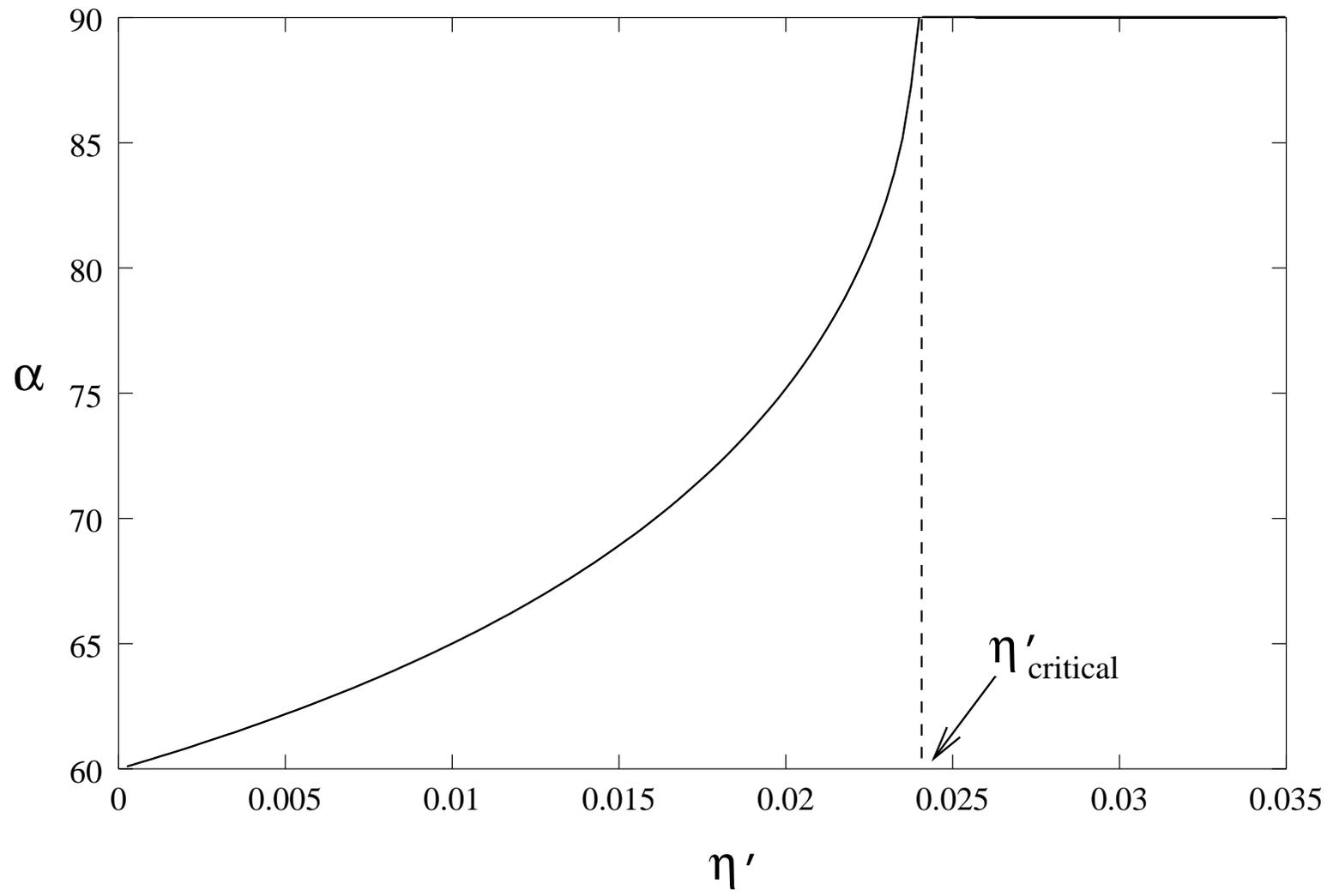
\end{document}